\documentclass[11pt]{article}
\pdfoutput=1
\usepackage{amsmath,epsf,amssymb,latexsym,enumerate,cite,shadow,array,color}
\usepackage{slashed}
\usepackage{graphicx,wasysym}

\DeclareFontFamily{U}{rsf}{} \DeclareFontShape{U}{rsf}{m}{n}{
  <5> <6> rsfs5 <7> <8> <9> rsfs7 <10-> rsfs10}{}
\DeclareMathAlphabet\Scr{U}{rsf}{m}{n} \makeatletter
\@addtoreset{equation}{section} \makeatother

\def\be{\begin{equation}}
\def\ee{\end{equation}}
\def\ba{\begin{array}}
\def\ea{\end{array}}

\newcommand{\beq}{\begin{equation}}
\newcommand{\eeq}[1]{\label{#1}\end{equation}}
\newcommand{\bea}{\begin{eqnarray}}
\newcommand{\eea}[1]{\label{#1}\end{eqnarray}}

\def\K{K{\"a}hler}
\def\vp{{\varphi}}

\usepackage{ifpdf}
\ifx\pdfoutput\undefined
   \pdffalse
\else
   \pdfoutput=1
   \pdftrue
  \usepackage[pdftex]{hyperref}
\newcommand{\rf}[1]{(\ref{#1})}

\begin{document}

\begin{titlepage}
\hfill CERN-TH/13-178\\

\hskip 1cm
\vskip 1cm
\begin{center}

{\huge \bf{Minimal Supergravity Models of Inflation  
}}

\vskip 0.8cm  

{\bf \large Sergio Ferrara$^{1,2}$, Renata Kallosh$^3$,  Andrei Linde$^3$ and Massimo Porrati$^4$}  

\vskip 0.5cm

\noindent $^{1}$ Physics Department, Theory Unit, CERN, CH 1211, Geneva 23, Switzerland\\
$^{2}$ INFN - Laboratori Nazionali di Frascati, Via Enrico Fermi 40, I-00044 Frascati, Italy~\footnote{On leave of absence from Department of Physics and Astronomy, University of California Los Angeles, CA 90095-1547 USA}\\
$^3$ Department of Physics and SITP, Stanford University, Stanford, California
94305 USA\\
$^{4}$ {CCPP, Department of Physics, NYU
4 Washington Pl. New York NY 10016, USA}

\end{center}

\vskip 1.5 cm

\begin{abstract}

We present a superconformal master action for a class of  supergravity models with one arbitrary function defining the Jordan frame. It leads to a gauge-invariant action for a real vector multiplet, which upon gauge fixing describes a massive vector multiplet, or to a dual formulation with a linear multiplet and a  massive tensor field. In both cases the models have one real scalar, the inflaton, naturally suited for single-field inflation.  Vectors and tensors required by supersymmetry to complement a single real scalar do not acquire vev's during inflation, so there is no need to stabilize the extra scalars which are always present in the theories with chiral matter multiplets. The new class of models can describe any inflaton potential which vanishes at its minimum and grows monotonically away from the minimum. In this class of supergravity models one can fit any desirable choice of inflationary parameters $n_{s}$ and $r$.

 \end{abstract}

\vspace{24pt}
\end{titlepage}


\tableofcontents

\newpage

\section{Introduction}

The recent cosmological observations by the Planck probe~\cite{Ade:2013rta}, beside supporting earlier findings of WMAP~\cite{Hinshaw:2012aka}, suggested that a single scalar field model of inflation might explain the data. 
This particular  feature, as well as the actual values of the cosmological observables $n_s$ and $r$ led to a burst  of activity recently. One would like to try to use this information as a hint towards the fundamental theory behind the observations. The goal of such a bottom up approach, initiated in~\cite{Kallosh:2010ug} a couple of years before the anticipated Planck results,  was to end up, upon moduli stabilization, with general potentials capable of describing any possible outcome of the future data. The philosophy was  to start with the analysis of such models in supergravity and eventually uplift this information to string theory and its landscape.    These developments, including the work of some of the present authors \cite{Kallosh:2013pby,Kallosh:2013lkr,Kallosh:2013hoa},  continued  after the Planck data were released.
 
The purpose of this paper is to add a new way of thinking about supergravity in view of recent cosmological data. 
We find here that the data from the sky cast in a new light some supergravity papers from 70's and 80's, see   \cite{Ferrara:1983dh,Kugo:1983mv,Cecotti:1987sa,Cecotti:1987qr,cfps,vp,Mukhi:1979wc,cfgvp} and references there. Our analysis here will be build on extension of the results obtained in those papers.

For example, in the past a supergravity version of the Starobinsky model $R+\gamma R^2$ \cite{Starobinsky:1980te}\footnote{The original model \cite{Starobinsky:1980te} was based on the Einstein theory with  conformal anomaly, but later it was modified and cast to its  more familiar form $R+\gamma R^2$ \cite{Kofman:1985aw}. When we will be talking about the potential in this model, we will have in mind its dual representation in terms of the Einstein gravity and a scalar field with the potential $V\sim (1-\exp(-\sqrt{2/3}\vp))^2$  \cite{Whitt:1984pd}.} was developed in the old-minimal formulation  in~\cite{Cecotti:1987sa} and in the new-minimal version of supergravity in~\cite{cfps}. When one tries to implement inflation in the simplest version of the model described in \cite{Cecotti:1987sa},  one finds a tachyonic Goldstino instability at large values of the inflaton field. This problem was recently resolved  in \cite{Kallosh:2013lkr,Kallosh:2013hoa}, where several different versions of a stable supersymmetric generalization of the theory $R+\gamma R^2$ exactly reproducing the inflaton potential of  \cite{Starobinsky:1980te} have been constructed.  Related approaches were proposed in \cite{Ketov:2010qz,Ellis:2013xoa,Buchmuller:2013zfa,Ellis:2013nxa,Farakos:2013cqa}. In particular, a stimulating proposal of \cite{Ellis:2013xoa} was to embed the Starobinsky model in the no-scale supergravity   \cite{Cremmer:1983bf}, which exactly reproduced the inflaton potential of  \cite{Starobinsky:1980te}   prior to the moduli stabilization.  However, moduli stabilization required breaking of the no-scale structure of the theory, which slightly modified the inflationary potential of \cite{Ellis:2013xoa}, see  \cite{Ellis:2013nxa}. In this respect, it is interesting that in the supergravity version of the theory $R+\gamma R^2$ developed in \cite{cfps} the issue of stabilization of the extra moduli does not appear at all because this theory does not contain any other scalars except  the inflaton \cite{Farakos:2013cqa}.

A major step forward in constructing inflationary
models in supergravity was made in \cite{Kawasaki:2000yn}, where an implementation of the simplest chaotic inflation model $m^{2}\phi^{2}$  \cite{Linde:1983gd}  was proposed in the supergravity context. 
This was significantly generalized in  \cite{Kallosh:2010ug}, where a general class of chaotic inflation models with a nearly arbitrary form of the inflaton potential was developed using chiral supergravity matter. To provide consistent  cosmological models, certain conditions on the \K\, potentials were required for stabilization of other moduli, which can be achieved by the methods developed in  \cite{Kallosh:2010ug}. The setting of this approach becomes now a part of the full supergravity landscape as new features are emerging. 

In this paper, we will show that the landscape of supergravity models capable of explaining the Planck data can be substantially extended. We  present a novel set of supergravity models were the physical multiplets are not chiral but vector or linear multiplets in the Higgs phase, where they are massive. We will show a ``master'' superconformal gauge-invariant model with a vector or a  linear multiplet both in the old-minimal as well as in the new-minimal formulation. The model exists in a Higgs phase with a massive vector or tensor, as well as in a de-Higgsed (ungauged) phase where the vectors and tensors are massless but there is a second scalar.

We end up with a cosmological model for a single scalar with the potential given by one nearly arbitrary function. There are two  constraints on the potential: it should vanish at its minimum, and it should monotonically grow away from the minimum. Both of these conditions can be relaxed by considering theories with many scalar fields, but from the point of view of inflation these constraints are not really restrictive: potentials of this type are quite generic and ideally suited for implementation of a single field chaotic inflation  \cite{Linde:1983gd}. By tuning the shape of the potential, one can fit {\it any}\, desirable values of the observational parameters $n_{s}$ and $r$ consistent with recent observational data.

On the other hand, in some cases one may encounter the cosmological attractor mechanism, when a broad class of different theories make very similar predictions for $n_{s}$ and $r$. An example of such mechanism was recently given  \cite{Kallosh:2013hoa} in a class of models using chiral superfields. In this paper, we will find a similar result for a certain class of models resembling the model $R+\gamma R^2$ in the theories with a vector multiplet. 

Our ``minimal'' class of  models with a single vector matter multiplet can be generalized by adding other chiral and vector multiplets as well as any superpotential that respects R symmetry, see for example  \cite{Ferrara:1988qxa}.
During inflation supersymmetry is broken since at that time the auxiliary field $D$ does not vanish. However, after inflation, when the inflaton field is at the minimum of its potential, $D=0$ in our minimal model  and supersymmetry is restored. Extension of our inflationary model to a realistic model with supersymmetry breaking requires separate investigation of the low energy scale susy breaking mechanisms, for example via hidden sectors and soft susy breaking terms, which should not affect the inflationary regime.

This paper is organized as follows:

In Section 2, we define the ``master" supergravity action with chiral compensator that reduces to supergravity coupled to either a massive vector multiplet or to a massive linear multiplet, upon integrating out appropriate non-dynamical superfields. 

In Section 3, we present  models which are dual to those in Sec. 2.  Here the master superconformal action is based on a linear compensator. In particular, we show how, as in~\cite{cfps}, a
special choice of this master action reproduces the supergravity generalization of the $R + \gamma R^2$ action in the new minimal formulation.

In Section 4 we present the simple part of these new supergravity models  relevant to inflation, where only scalars have non-vanishing vet's. The action thus depends on a single scalar $C$, the first component of the vector superfield, which has a non-canonical kinetic term. We rewrite this action in terms of a canonically normalized field $\vp$ and explain the relation between these two versions of the model. We present as examples the  simplest model of chaotic inflation ${m^{2}\over 2}\vp^{2}$  \cite{Linde:1983gd}, more general chaotic inflation models, and the T-model ~\cite{Kallosh:2013hoa}. We also present a class of  $SU(1,1)/U(1)$ gauged sigma models,  which contains, as  particular cases, the supersymmetric version of the theory $R+\gamma R^2$ developed in~\cite{cfps}, and its deformations corresponding to exponents of the kind $e^{-b \vp}$ where $b$ is arbitrary instead of being equal to $\sqrt {2/3}$.  Other interesting  inflationary potentials related to  brane supersymmetry breaking and integrable systems have  been studied in \cite{Fre:2013vza}.

In Appendix A we summarize the result of~\cite{cfps}, which shows the equivalence of new minimal $R+\gamma R^2$ gravity with a massive vector multiplet. In Appendix B, we give an explicit component proof of the equivalence between the (bosonic part of) a massive vector Lagrangian and a massive tensor Lagrangian.

We are using structures and notation of  (CFPS) paper \cite{cfps}.
Readers interested only in cosmological applications may start with Section 4, with understanding that  various embeddings of eq. \rf{m41} into  supergravity with massive vectors or tensors are explained in earlier sections.

\section{Superconformal Master Model with Chiral Compensator }
The superconformal ``master'' model  depends on two real vector 
multiplets, $U$ and $V$, one linear multiplet $L$ and a chiral conformal compensator $S_0$,
\beq
{\cal L} (S_0, U, L, V; g)=- S_0 \bar S_0 \Phi(U)_D +L (U-gV)_D+{1\over 2 } [ W_\alpha (V) W^\alpha(V) +h.c.]_F .
\eeq{m1}
The conformal/chiral weights of these superfields are the following: for $U$ and $V$ we have $w=n=0$, for $L$ we have $w=2, n=0$.  For $S_0$ we have $w=1, n=1$ and for $\bar S_0$ we have $w=1, n=-1$. There is a single dimensionless parameter $g$, a coupling between the linear and vector multiplets. 

Our master model depends on a real function $\Phi$ of a real vector multiplet. This function has to be strictly 
 positive,  so that it describes gravity, rather than anti-gravity. For example, pure supergravity is the case $\Phi(U)=1$
We will also need the following relations between superfields
\beq
\Phi = e^{-{1\over 3} {\cal J}} \, , \qquad    \, {\cal J}\equiv  -3 \log \Phi  .
\eeq{m3}
\subsection{Physical vector multiplet  (massive or massless vector field)}

By varying the linear multiplet in action \rf{m1} we find that  $U= \Lambda +\bar \Lambda + gV$ with $\Lambda$ chiral. We find that our master action becomes
\beq
{\cal L}= -S_0 \bar{S}_0 \Phi(\Lambda+\bar{\Lambda} + g V)_D + {1\over 2 } [W_\alpha(V) W^\alpha(V) + h.c]_F .
\eeq{m4}
The action has both superconformal symmetry and a gauge symmetry 
\beq
\Lambda \rightarrow \Lambda + g\Sigma, \qquad V \rightarrow V -\Sigma -\bar{\Sigma},
\eeq{m4'}
where $\Sigma$ is a chiral superfield. 

Note that by defining $S\equiv\exp(\Lambda)$ and   $\Phi(U)\equiv f (Se^{g V}\bar{S})$, this is a particular case of the standard supergravity Lagrangian~\cite{cfgvp}. This special form will play a most important role in the cosmological applications of the model.  Notice also that ${\cal J}(\Lambda + \bar{\Lambda})$ describes a K\"ahlerian sigma model invariant under a (gauged) shift symmetry. 

 We may decouple the Abelian  vector multiplet $V$ from supergravity by taking the limit $g\rightarrow 0$. In such case we recover a model of a free massless vector multiplet from the second term of our superconformal action. From the first term we recover instead a model of supergravity interacting with a chiral multiplet 
 without a superpotential, but with a generic \K\, potential, depending on $\Phi (\Lambda +\bar{\Lambda})$.

If however, $g\neq 0$ ,we can use the Abelian gauge symmetry  \rf{m4'} to fix the gauge symmetry by requiring
\beq
\Lambda=0.
\eeq{m5}
Then the Lagrangian becomes that of a massive vector field:
\beq
{\cal L}=-S_0 \bar{S}_0 \Phi (gV)_D + [W_\alpha(V) W^\alpha(V)+ h.c. ]_F .
\eeq{m6}
We  can also gauge-fix the superconformal symmetry by requiring that 
\beq  
S_0 = M_{P}= {1}
\eeq{m7}
for the Weyl weight 1 superfield $S_0$. For $g=1$ this is the    self-interacting massive vector multiplet action \cite{vp} in the Jordan frame. One can either perform the change of variables, starting with this action and find the action in the Einstein frame, as was done in  \cite{vp}  or alternatively, use the different super-conformal superfield gauge  
\be
S_0 \bar{S}_0 \Phi (gV)_D=1
\label{SC}\ee
which will lead directly to an Einstein frame action. 
In this way,  the bosonic part of our superfield action in the gauge \rf{SC} corresponds to the  bosonic action in \cite{vp} where the gauge coupling $g$ is now restored \footnote{We note that eq. \rf{m10} has the correct normalization, since the vector and scalar square masses are both equal to $m^2 =- g^2 J''$ at the supersymmetric minimum $J'=0$. The potential in \rf{m15} becomes the same as in \cite{Farakos:2013cqa} by sending $g\rightarrow 2g$ and 
$\vp \rightarrow \sqrt 2 \vp$.}
\beq
{\cal L}= -{1\over 2}R -{1\over 4} F_{\mu\nu}(B)F^{\mu\nu}(B) + {g^2\over 2} J'' B_\mu B^\mu
+{1\over 2} J'' (C)\partial_\mu C\partial^\mu C -{g^2\over 2} J'^2(C).
\eeq{m10}
Here $C$ is the scalar in the vector multiplet, $B_\mu$ is its vector and $'$ is differentiation w.r.t. $C$.
Our superfield action \rf{m1} was defined for an arbitrary function $\Phi= e^{-{1\over 3} {\cal J}}$. The arbitrary function $J(C)$ in the component action in \rf{m10} is related to a superfield one as follows
\beq
J= -{1\over 2} {\cal J} +\rm {const}= {3\over 2} \log \Phi  +\rm{const} .
\eeq{m10'}
Note that the  Einstein frame  was obtained in~\cite{vp} after conformal rescaling  with the factor $e^{{1\over 3} J}$ on the vierbein, and on other fields accordingly, starting from the Jordan frame action. 

A particular example of $\Phi$ was described in detail in \cite{cfps}. 
There it was found there that a particular choice of  the \K\,  potential reproduces the pure $R+\gamma R^2$ theory in new minimal  supergravity. The derivation is reviewed in Appendix A. 
 The specific potential is 
 \beq
 \Phi=-C\exp C\, , \quad J={3\over 2} \bigl[\log (-C) + C\bigr]  , \quad J'={3\over 2}(C^{-1} +1)\, ,\quad J''=-{3\over 2}C^{-2} .
 \eeq{m14}
 Notice that the \K\, potential gives a manifold $SU(1,1)/U(1)$, where the gauged symmetry is the translational isometry. 
 To canonically normalize the scalar one needs $C=-\exp(\sqrt{2/3}\vp)$ so that the scalar potential becomes that of the Starobinsky model \cite{Starobinsky:1980te}, see also a more recent paper on this \cite{Farakos:2013cqa}
\beq
V={9\over 8}g^2\bigl[1-\exp(-\sqrt{2/3}\vp)\bigr]^2.
\eeq{m15}
In our class of models, the \K\, potential of the  $SU(1,1)/U(1)$ manifold leads to the D-term potential \rf{m15}. Our models differ from those with
chiral multiplets and F-term potentials. 
The presence of the vector field in our models is a condition for the existence of the D-term potential since the $D$-field is an auxiliary field of the vector multiplet.   Vector fields typically do not play any role during inflation. However, at the exit from inflation during reheating and creation of matter the presence of the vector field might be important and needs  separate study.

Note that the D-term potential ${g^2\over 2} J'^2(C)
$ originates from the real auxiliary field $D$ of the vector multiplet, where
\beq
D(C)= J'(C),
\eeq{m16}
and that the positivity of the moduli space metric of the scalar $C$ requires that
\beq
J''(C) <0 \ , 
\eeq{m17}
which is satisfied by the function in equation (\ref{m14}).

\subsection{De-Higgsing, $g=0$ limit}

Our master model has two phases: one, the Higgs phase,  with a massive vector and a real scalar at $g\neq 0$ ; the other, the de-Higgsed phase,  with one massless vector and one complex scalar, with $g=0$.
To retrieve the component form  of our action \rf{m4}, which allows both phases, we have to define $A_\mu = B_\mu +{1\over g} \partial_\mu a$,  so that 
\beq
{g^2\over 2} J'' B_\mu B^\mu = {g^2\over 2} J''(A_\mu + {1\over g} \partial_\mu a)^2. 
\eeq{m12}
This explains how a massive vector ``eats an axion.'' In the limit $g\rightarrow 0$ this term leaves us with a kinetic term for the axion
\beq
{g^2\over 2} J'' B_\mu B^\mu \rightarrow  {1\over 2} J''(  \partial_\mu a)^2.
\eeq{m13}
It restores the de-Higgsed phase of the model, where in components we get
\beq
{\cal L}= -{1\over 2}R -{1\over 4} F_{\mu\nu}(A)F^{\mu\nu}(A) +  {1\over 2} J''(  \partial_\mu a)^2
+{1\over 2} J'' (C)\partial_\mu C\partial^\mu C ,
\eeq{m10''}
i.e. a massless vector, 2 uncharged scalars and no potential. Thus, the Lagrangian \rf{m10} at $g\rightarrow 0$ becomes a $\sigma$-model Lagrangian of a complex scalar $\Lambda|_{\theta=0}\equiv z=C+ia$ with \K\, potential 
$-{1\over 2}J({z+\bar z\over 2})$. 

This is in agreement with our analysis of the superfield action when at $g\rightarrow 0$ the \K\, potential depends on ${\Lambda+\bar \Lambda}$.  This is also in agreement with  \cite{cfps}. 

\subsection{Physical linear multiplet (massive or massless tensor field) }
We start again with the master action~(\ref{m1}), differentiate it with respect to $U$ and solve for $L$ to get the dual linear multiplet action
\beq
 {\cal L}(S_0,L,V,g)=\left[S_0 \bar S_0 F \Big ({L\over S_0 \Bar S_0}\Big )- g L  V\right]_D+{1\over 2 } \bigl[W_\alpha (V) W^\alpha(V) +h.c.\bigr]_F\ ,
\eeq{m21}
with
\beq
F(U)= U\Phi'(U)-\Phi(U) \mbox{ computed at } \Phi'(U)= {L\over S_0 \Bar S_0}.
\eeq{m20}

Lagrangian~(\ref{m21}) is in reality independent of $V$. To show this we recall that
the linear multiplet is constrained: it can be expressed via a chiral spinor superfield $L_\alpha$ as:
\beq
L=( D^\alpha L_\alpha +\bar D_{\dot \alpha} \Bar L^{\dot \alpha}).
\eeq{m20'}
Therefore we may integrate by part the term $-[gLV]_D$ to obtain:
\beq
 -[g L  V ]_D= g [ L^\alpha W_\alpha + \bar L_{\dot \alpha} \bar W^{\dot \alpha}]_F .
 \eeq{m22}
The action depends on the unconstrained field $L_\alpha$ only via $L$, therefore there is a gauge symmetry
\beq
L_\alpha  \rightarrow L_\alpha -{i\over g}  W_\alpha \, , \mbox{  if  }D_\alpha W^\alpha = \bar D_{\dot \alpha} \bar  W^{\dot \alpha}.
 \eeq{m2}
 The action contains the following $V$-dependent terms
 \beq
 {1\over 2} [W^\alpha (V) W_\alpha (V) + g W^\alpha(V) L_\alpha + h.c. ]_F.
 \eeq{ms1}
 By introducing a chiral Lagrange multiplier $M_\alpha$, this F term containing $V$ can be rewritten as
 \beq
 -{1\over 2} [M^\alpha M_\alpha + M^\alpha W_\alpha(V) + i g M^\alpha L_\alpha + h.c.]_F.
 \eeq{ms2}
 So, by integrating out $M^\alpha$ and using the gauge symmetry~(\ref{m2}), the action becomes $V$ independent:
\beq
 {\cal L}(S_0,L^\alpha, g)= S_0 \bar S_0 F \Big ({L\over S_0 \Bar S_0}\Big ) - {g^2\over 2} (L^\alpha L_\alpha +h.c.).
\eeq{m24}
After gauge fixing the conformal compensator, this action  produces in components the following massive tensor field action 
\beq
{\cal L}= -{1\over 2}R - {1\over 2} (J'')^{-1}  (\partial_{[\mu} B_{\rho \sigma] }) ^2-{1\over 4} g^2 B_{\rho \sigma}^2
+{1\over 2} J'' (C)\partial_\mu C\partial^\mu C -{g^2\over 2} J'^2(C).
\eeq{m25}

This bosonic Lagrangian is equivalent to eq.~(\ref{m10}) as it was shown e.g. 
in~\cite{Cecotti:1987qr} and reviewed in Appendix B.

\section{Superconformal Master Model with Linear Compensator }
In principle, all the Lagrangians given in the previous section can be converted into a ``new minimal" auxiliary field form since, in the physical linear multiplet language, the mass term $gLV$ does not depend on the compensator. In particular, the ``master action" can be written in the new minimal formulation simply as 
\beq
{\cal L}(L_0, L,U,V,g)= \bigl[L_0 \log( L_0/S_0\bar{S}_0) + {1\over 3} L_0 {\cal J}(U) +
L(U-gV)\bigr]_D +{1\over 2} \bigl[W^\alpha(V)W_\alpha (V) +h.c.\bigr]_F .
\eeq{ms3}

By variation of $L$ we recover both the K\"ahlerian gauged sigma model in the new minimal formulation, as found in~\cite{cfps} eq. (1.7), and the massive vector model --which is the former in the unitary gauge. By varying with respect to $U$ we recover the new minimal form of the massive linear multiplet Lagrangian:
\beq
{\cal L}(L_0,L^\alpha, g )= [L_0 \log( L_0/S_0\bar{S}_0) + L_0 M(L/L_0)]_D -{1\over 2}[g^2 L^\alpha L_\alpha + h.c. ]_F,
\eeq{ms4}
where 
\beq
M(L/L_0)= {1\over 3} \left[{\cal J}(U) - U {\partial {\cal J}\over \partial U}\right] \mbox{ computed at } {1\over 3}{\partial {\cal J}\over \partial U}=-L/L_0.
\eeq{ms5}

In the tensor multiplet formulation, the scalar potential term comes from the mass term of the linear multiplet scalar, $\sigma$, which is the first component of the supermultiplet $L/L_0$. From eq.~(\ref{ms5}) we notice that $\sigma=(2/3)J'(C)$. This explains why the potential is proportional to $g^2 J'^2$.

For the choice of function ${\cal J}/3=-\alpha \log U - \beta U$ --which reduces to the Starobinsky model given in eq.~(\ref{m14}) for $\alpha=\beta=1$--  the Lagrangian~(\ref{ms4})  acquire a particularly simple form. Written in terms of $V$ and $L$ it reads:
\beq
{\cal L}=[(1-\alpha)L_0\log L_0 +\alpha L_0 \log[(L-\beta L_0)/S\bar{S}_0] -g L  V]_D+{1\over 2 } [W_\alpha (V) W^\alpha(V) +h.c.]_F.
\eeq{ms5a}

It is now immediate to show that the $R+\gamma R^2$ new minimal supergravity can be recovered from eq.~(\ref{ms5a}) for $\alpha=1$ and $\beta\neq 0$ arbitrary.  By varying  w.r.t. $X\equiv L_0+L/\beta $  one can solve for the vector superfield as $g\beta V_L=\log(L_0-L/\beta) + \mbox{chiral } +\mbox{antichiral}$. By substituting into~(\ref{ms5a}) one obtains the Lagrangian~\cite{cfps}
\bea
{\cal L}(L_0-L/\beta)&=&[S_0 \bar{S}_0 V_Le^{V_L}]_D + {1\over 2g^2\beta^2} [W_\alpha (V_L)W^\alpha(V_L) + h.c. ]_F, \\ 
&& V_L\equiv \log[ (L_0-L/\beta)/S_0\bar{S}_0], \qquad W_\alpha=\Sigma D_\alpha V_L.
\eea{ms5aaa}
Here $\Sigma$ is the chiral projector defined in \cite{Kugo:1983mv} and the Lagrangian we obtained is identical with that of~\cite{cfps} (cfr. Appendix A) upon redefining 
$L_0-L/\beta \rightarrow L_0$, $1/g^2\beta^2 \rightarrow \gamma $. 

\section{From  Supergravity with Massive Vector/Tensor to Inflation}
During inflation  in the FLRW universe there is no vev of the vector or tensor fields, so we need only the scalar-gravity part of the action  \rf{m7} or  \rf{m25}.
\beq
e^{-1} L =  -  \frac{1}{2}R + \frac{1}{2}  J'' (\partial_\mu C )^2 -  \frac{g^2}{2} (J ' )^2 \, , \qquad J''(C) <0 \ ,
\eeq{m41}
where $'$ denotes as before differentiation over the single real scalar $C$. The scalar potential has an extremum at
\be
V'=g^2 J'' J'=0 \ .
\ee
Since $J''$ must be non-vanishing and negative we find that 
\be\label{nomax}
V'= g^2J'' J'=0\, ,\qquad \Rightarrow J'=D=0\, \qquad V''= g^2[(J'')^2 + J''' J']|_{J'=0}= g^2(J'')^2>0 \ ,
\ee
and supersymmetry is restored at the minimum of the potential. It is interesting that the potential in this theory vanishes at its minimum and it  must grow monotonically away from this minimum in the domain of stability of the theory. Indeed, if the potential grows and then  starts decreasing, this involves the change of sign of $D' = J''$, which would imply a wrong sign of the kinetic term of the field $C$, and, consequently, vacuum instability.

We can rewrite the action using a  canonical  field $\vp$ instead of a non-canonical  $C$
and change variables from $C$ to $\vp$. Let us introduce some definitions
\be
D (C)\equiv J' (C)\ , \qquad  D'(C)\equiv  J''(C) \ .
\ee
After a change of variables the action is
\begin{equation}\label{chaotmodel}
e^{-1} L =  - \frac{1}{2}R - \frac{1}{2}(\partial_\mu \varphi)^2 - \frac{g^2}{2} (D (\vp))^2 \ ,
\end{equation}
where 
\be
D (C(\vp))\equiv  D(\vp) \ .
\ee
and 
\be
\left({d\vp\over dC}\right)^{2} = {-D'(C) }\ .
\ee
From the relation 
\be
D'(C) = {d D\over d \vp}\, {d\vp\over dC}
\ee
one finds
\be
\Big | {d D\over d \vp}\Big | = \sqrt { -D'(C)} \ .
\ee
Thus
\be\label{relation}
{d D\over d\vp}= - {d\vp\over dC} \ .
\ee
Note also that the vector mass is always equal to $|g \, {dD\over d\vp}|$, see (\ref{m10}).
At the supersymmetric minimum $D=0$, it equals the inflaton mass, as it should. 

The rest of the fully supersymmetric action action requires various $C$-dependent terms like $J'$,  $J''$, $J'''$, $J''''$, which become $\vp$ dependent terms. We may define the model completely either by giving $J'(C)$ and computing higher derivatives to specify the full eq. (3.9) in \cite{vp},  or we may codify each model by the choice of $D(\vp)$. In such case we may find the complete action first by defining the relation between these variables 
\be\label{cphi}
C(\vp)= - \int d\vp \Big ({d D(\vp) \over d \vp}\Big )^{-1} \ ,
\ee
and looking for the inverse one, $\vp (C)$, which should allow us to find 
 $ J' (C)=D (\vp(C))$ for any given $D(\vp)$.

The generic case of \rf{chaotmodel} has a positive potential which is the square of an arbitrary function $D (\vp)$,
 and therefore offers a possibility to fit any value of the cosmological observables as long as  that  potential has a slow-roll regime. In all previous models of inflation in supergravity  the issue of stabilization of remaining moduli required a significant effort, see for example \cite{Kallosh:2010ug}, as well as many other papers on this.
 
 This situation would be perfect sometime ago, but now that we have to fit the data from Planck.  There is  one obvious perfect feature of this model: it has exactly one  scalar inflaton, and there is nothing to stabilize despite having a complete supergravity model. Plank data agree with a single scalar inflaton model, so let us take it as a starting point for embedding inflation in supergravity.
 
 \subsection{The simplest chaotic inflation model ${g^{2}\over 2}\vp^{2}$.}

Now we are going to investigate the possibility to embed some  well known inflationary models in the theory discussed above.  The main problem is to make sure that these inflationary potentials can be cast in the form $D^{2}(C)/2$ with $D'<0$ as required in this theory. As we will see, this can be done for a very broad class of inflationary potentials, but some non-trivial restrictions do appear.
 
 We will begin with the simplest chaotic inflation model of a canonically normalized scalar field $\vp$ with the quadratic potential
 \be\label{quadr}
 V(\vp) = {g^{2}\over 2}\vp^{2} \ .
 \ee
 This potential can be represented as $g^2 D^{2}(\vp)/2$, where
\be
D(\vp) = - \vp
\ee
Then ${d D\over d \vp}=   -1$,  and therefore
\be
C=  \int d\vp =  \vp \ .
\ee
Therefore $D(\vp)= -   C$ so that the condition $J'' = D'<0$ is satisfied, 
and the potential is given by (\ref{quadr}).

 This potential corresponds to the  simplest version of chaotic inflation \cite{Linde:1983gd}, with  the inflaton mass $m$ identified with $g$.  It played a very important role in the history of the development of inflationary cosmology, so the possibility to obtain this potential in such a simple way in supergravity is noteworthy. Interestingly, a supergravity model of this type, with a quadratic potential, was proposed back in 1979 \cite{Mukhi:1979wc}, but apparently its usefulness for cosmology was not appreciated by the very few people who were aware of its existence.
As we will see now, the theory which we are discussing in this paper can easily incorporate a much more general variety of inflationary potentials.

\subsection{Generic models of chaotic inflation}
The  model ${g^{2}\over 2}\vp^{2}$ described above represents the simplest version of chaotic inflation, but it is only marginally compatible with the Planck probe data, so we will try to generalize it now. Consider the following polynomial potential which can be represented as $g^2 D^{2}(\vp)/2$:
\be\label{three}
V(\vp) = {g^{2}\vp^{2}\over 2} (1-a\vp+b\vp^{2})^{2} \ .
\ee
This potential is polynomial and positive definite. It allows chaotic inflation for any values of its parameters  \cite{Linde:1983gd}. Observational data provide 3 main data points: The amplitude of the perturbations $\Delta_{R}$, the slope of the spectrum $n_{s}$ and the ratio of tensor to scalar perturbations $r$. Tensor perturbations have not been found yet, so we are talking about the upper bound $r \lesssim 0.1$. 

The potential  contains exactly 3 parameters which are required to fit these data, so we are not talking about fine-tuning where a special combination of many parameters is required to account for explaining a small number of data points; we are trying to fit 3 data points by a proper choice of 3 parameters, $g$, $a$ and $b$. The values of $n_{s}$ and $r$ do not depend on the overall scale of $V$; they are fully controlled by the parameters $a$ and $b$. One can show that by fixing a proper combination of $a$ and $b$  with a few percent accuracy, one can cover the main part of the area in the $n_{s} - r$ plane allowed by observations. After fixing these two parameters, one can find the proper value of $g \sim 10^{{-5}}$ to fit the observed value of $\Delta_{R} \sim 10^{{-5}}$. We will return to a detailed discussion of this issue in a separate publication; for a discussion of  $n_{s}$ and $r$ for very similar potentials in non-supersymmetric models see \cite{Destri:2007pv}. At present, for illustrative purposes, we will consider a particular set of parameters $a = 0.1$, $b = 0.0035$. This will help us to describe our general strategy by presenting some explicit potentials.

\begin{figure}[ht!]
\centering
\vskip 0.2cm \includegraphics[scale=0.6]{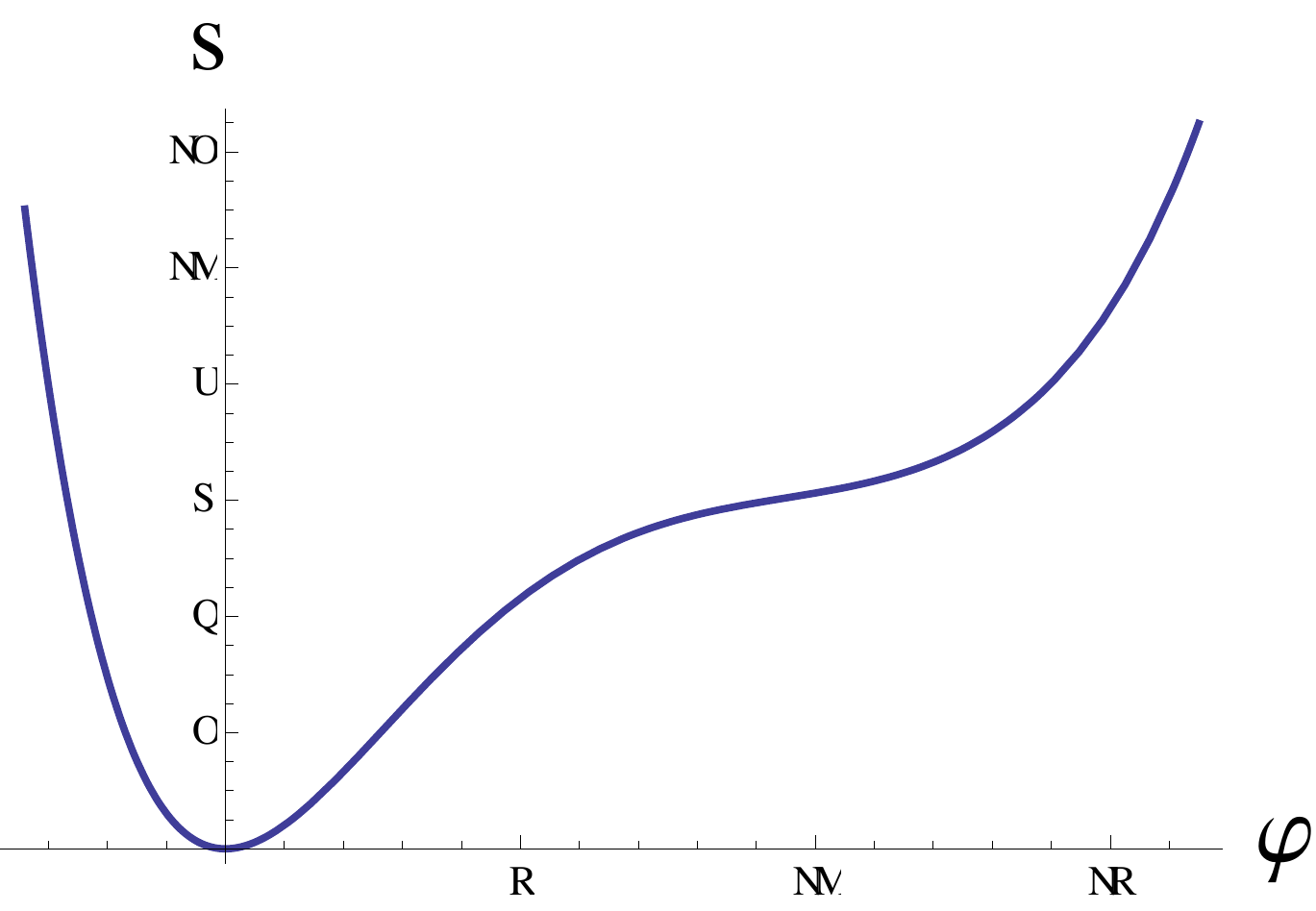}
\caption{Inflationary potential ${g^{2}\vp^{2}\over 2} (1-a\vp+b\vp^{2})^2$  (\ref{three}), for $a = 0.1$, $b = 0.0035$. The field is shown in Planck units, the potential $V$ is shown in units $g^{2}$. In realistic models of that type, $g \sim 10^{-5} - 10^{-6}$ in Planck units, depending on details of the theory, so the height of the potential in this figure is about $10^{-10}$ in Planck units.}
\label{potential}
\end{figure}

\begin{figure}[ht!]
\centering
\vskip 0.2cm \includegraphics[scale=0.4]{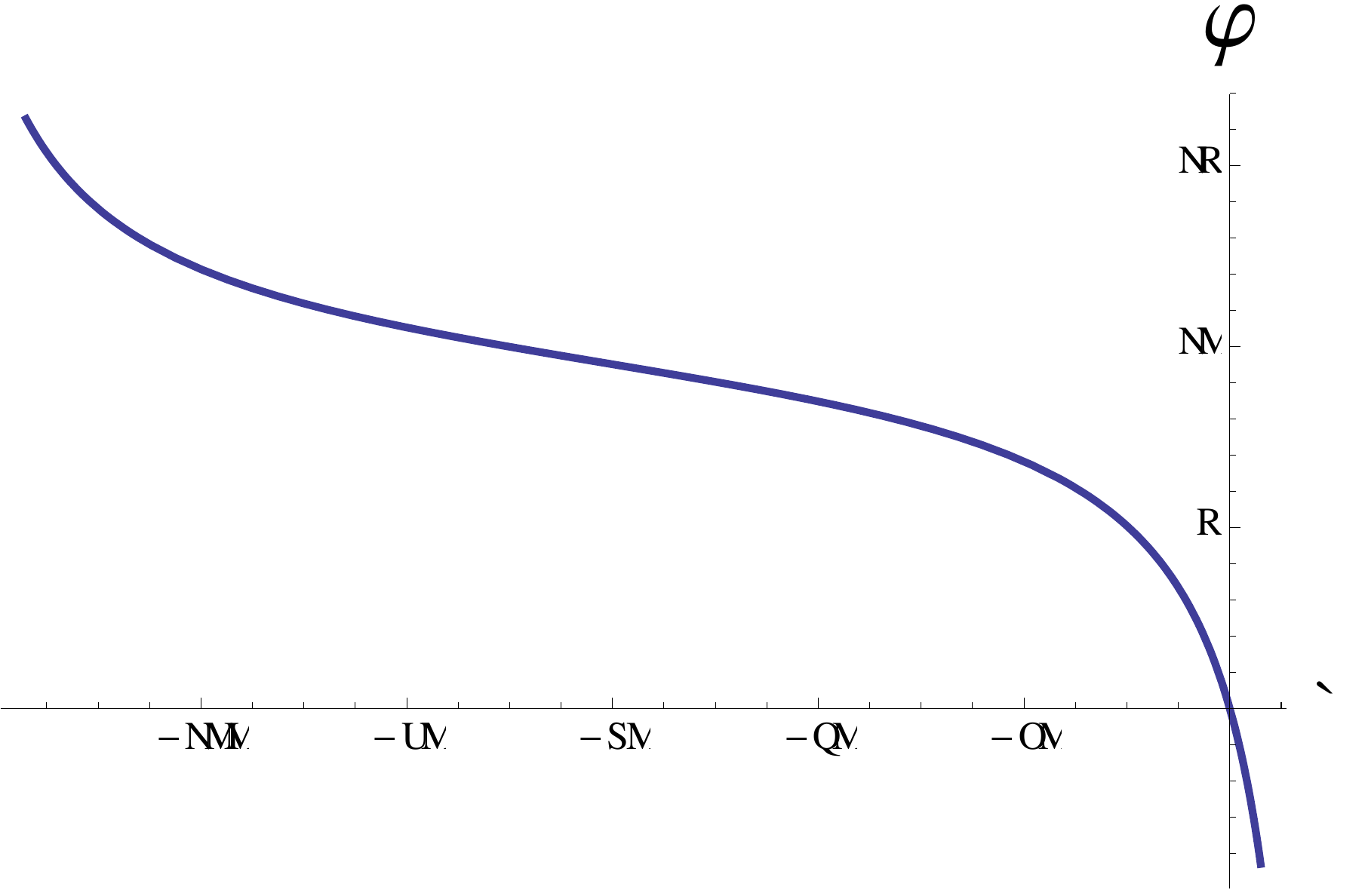}
\caption{Relation between the canonical variable $\vp$ and the original variable $C$ in the theory (\ref{three}). }
\label{2}
\end{figure}

\begin{figure}[ht!]
\centering
\vskip 0.2cm \includegraphics[scale=0.4]{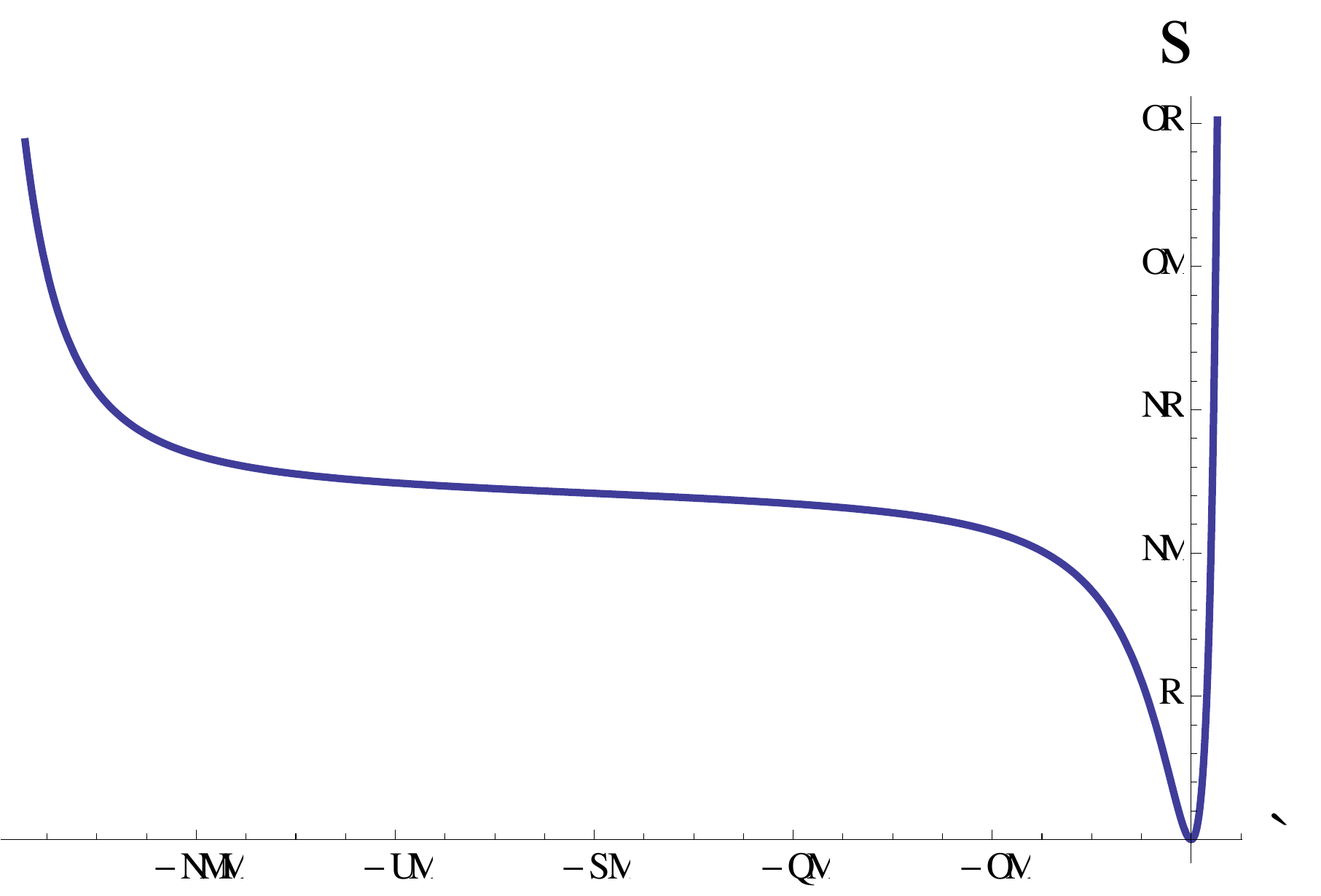}
\caption{Inflationary potential ${g^{2}\vp^{2}\over 2} (1-a\vp+b\vp^{2})^2$  (\ref{three}) as a function of the field $C$. }
\label{3}
\end{figure}

\begin{figure}[ht!]
\centering
\vskip 0.2cm \includegraphics[scale=0.4]{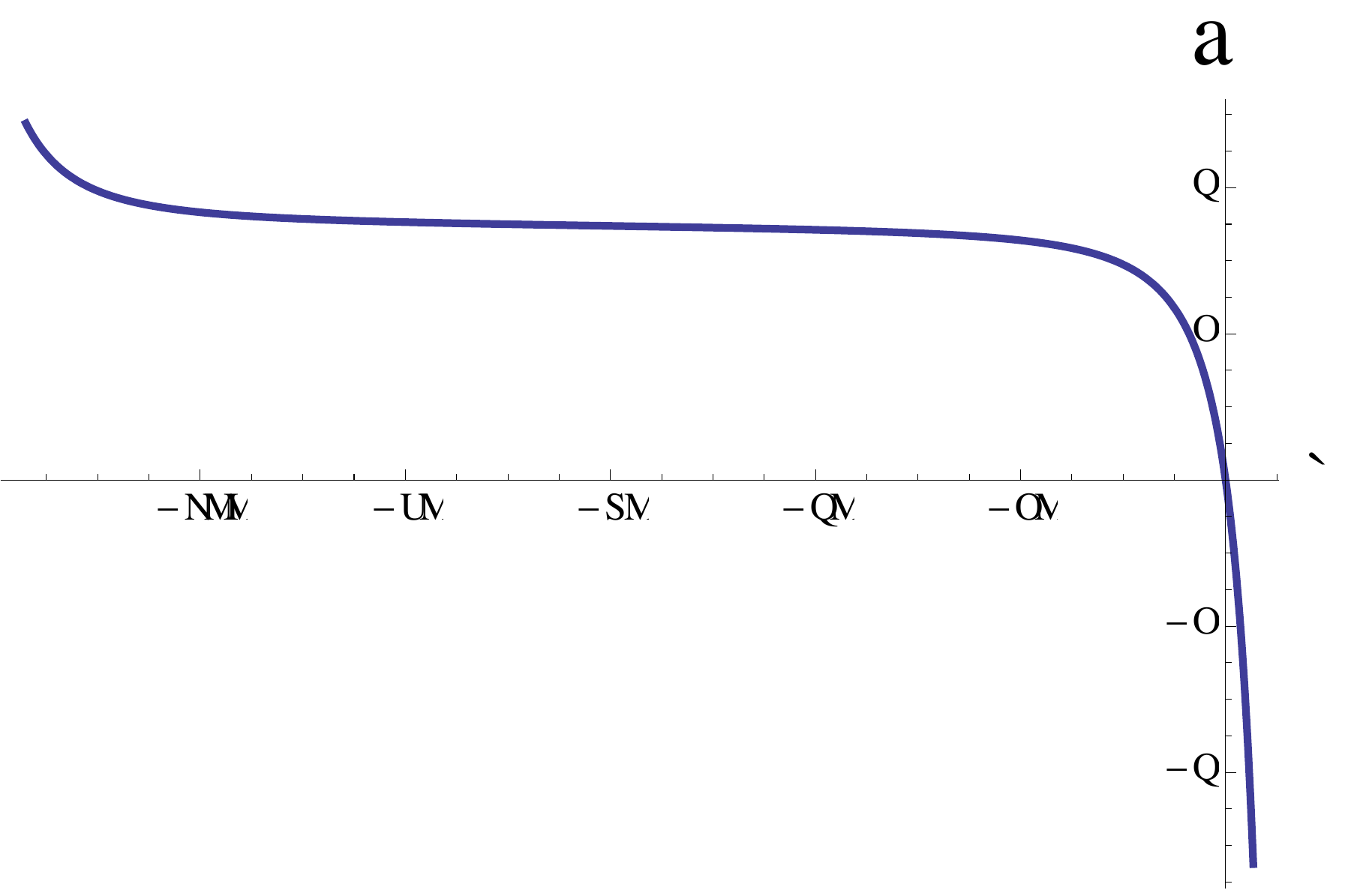}
\caption{The function $D(C)$. As we see, $D' <0$, so the condition $D'(C)<0$ is satisfied. }
\label{4}
\end{figure}

Our goal is to show that potentials of this type can be a part of our supergravity model. The potential (\ref{three}) can be represented as $D^{2}(\vp)/2$, with (note the signs!)
\be
D(\vp) = - \vp (1-a\vp+b\vp^{2}) \ , \qquad {dD\over d\vp} = - (1-2a\vp+3b\vp^{2}) \ ,
\ee
but we want to check that one can consistently express it as a function $D(C(\vp))$ satisfying the condition $D'<0$, where the derivative is taken with respect to $C$. For some inflationary potentials this can be done explicitly, using (\ref{cphi}) and finding an inverse function, whereas for some other models one should use numerical tools. This is a legitimate approach since one can study of inflationary consequences directly in terms of the canonical variable $\vp$, as long as we know that the corresponding functions have properties consistent with our general requirements. The basic idea is that instead of performing integration in (\ref{cphi}) and finding an inverse function, which may be complicated, one can numerically solve the differential equation for the function $\vp(C)$ thus finding the inverse function numerically. The corresponding equation is 
\be
{d\vp\over dC} = {dD\over d\vp} \ ,
\ee
where ${dD\over d\vp}$ should be calculated at the as yet undetermined $\vp(C)$. For example, in our case
\be
{d\vp\over dC} = - (1-2a\vp(C)+3b\vp^{2}(C))\ .
\ee
The results of our investigation are presented in Figs. \ref{2}, \ref{3}, \ref{4}, which show the functions $\vp(C)$, $D(C)$, and $V(C) = D^{2}(C)/2$. The relation between $\vp$ and $C$ is determined up to a constant. We have chosen it in such a way as to have the minimum of the potential at $C = 0$, similar to what we had in terms of $\vp$. As we see from the figures, the required condition $D'(C)<0$ is satisfied.

The procedure employed above is applicable for a broad class of theories where the potential has a minimum with $V = D^{2}/2 = 0$, and which grows monotonically away from the minimum. Indeed, the only real constraint of the class of admissible functions is that $D' < 0$ in the domain of stability of the inflationary model. In the models where $D(C)$ continuously decreases and passes through its zero, where the potential vanishes, this requirement is easily satisfied. In these models the potential, which is proportional to $D^{2}$, decreases towards the minimum, and then grows again, i.e. it grows when one moves away from the minimum in either direction. 

However, at the maximum of the potential $V = g^2 D^{2}/2$, the function $J'' = D'$  vanishes, which leads to  vanishing of the kinetic term of the field $C$, and then, with the change of sign of $D'$, the kinetic term changes its sign, which leads to ghosts and vacuum instability in the part of the moduli space beyond the maximum of the potential. This makes description of inflation at or beyond the maximum of the  potential problematic. This affects  models with metastable minima of the inflaton field such as old inflation, and also new, natural and hilltop inflation directly at the top or beyond the top of the potential. However, the admissible class of functions $D(C)$ is ideally suited for description of chaotic  inflation with potentials monotonically growing in both directions away from the minimum with $V = 0$. All models considered in this section satisfy this condition since the potential in each of them has {\it only one extremum}, the supersymmetric minimum at $D = 0$. When the field $C$ (or $\vp$) reaches the minimum, the function $D$ continues changing monotonically, so one can have $D' < 0$ all the way.

To summarize, our class of models is suitable for description of any inflationary potential which can be written as a square of a monotonically changing function $D(C)$ (or $D(\vp)$), which vanishes at some point, which correspond to the minimum of $V = D^{2}/2$. (Functions  $D(C)$ which never vanish are also possible, but in such models the potential does not have any minimum; such potentials are not necessarily useful for inflation, but they may describe dark energy.)
This class of functions is very general. It is just slightly more narrow than the class of functions which is allowed in the inflationary theories based on supergravity with chiral matter superfields  \cite{Kallosh:2010ug}. For example, according to   \cite{Kallosh:2010ug}, the inflaton potential can be given by $|f^{2}(\vp)|$, where $\vp$ is a real part of the scalar field. Here $f(\vp)$ is an arbitrary real holomorphic function of the field $\vp$. The corresponding potential can have any number of minima and maxima, unlike the potential discussed in this paper. However, the existence of many maxima and minima is not very helpful for describing observational data unless we study effects of tunneling at $N \sim 60$. In this sense the models discussed in the present paper are minimal, not requiring the existence of many scalars, while allowing us full functional freedom in tuning $n_{s}$ and $r$.

\subsection{Universality class of conformal inflation models, including the T-model,  the Starobinsky model, and  their deformations}

 In \cite{Kallosh:2013hoa} a new broad class of inflationary theories with spontaneously broken superconformal symmetry was found. These theories include various potentials  which asymptotically look like 
\be\label{expansion}
V(\varphi) = V_{*} \left[1 - \sum  a_{n} e^{- \sqrt{2/3}\, n\, \varphi} \right]
\ee
for $\vp > 0$ and  lead to inflation at  $\vp \gg 1$, or similar potentials $V_{*} \left[1 - \sum  a_{n} e^{+ \sqrt{2/3}\, n\, \varphi} \right]$ for $\vp <0$.  The {\it attractor property} of these models discovered in \cite{Kallosh:2013hoa} is the following: the observational predictions of all these models with arbitrary choice of $V_*$ and $a_n$ is the same, namely all these models predict that in the first approximation in $1/N$:
 \be
n_{s} =1 -2/N \approx 0.967 , \qquad r = 12/N^{2}  \approx  3.2 \cdot 10^{{-3}} \ .
\label{attractor} 
\ee 
One should note that these numerical estimates depend on the exact value of $N$, which in turn depends e.g. on the details of the post-inflationary dynamics, including physics of reheating. As a result, the exact results for $n_{s}$ and $r$ may slightly differ from the estimates given above for $N = 60$.

\begin{figure}[ht!]
\centering
\vskip 0.2cm \includegraphics[scale=0.55]{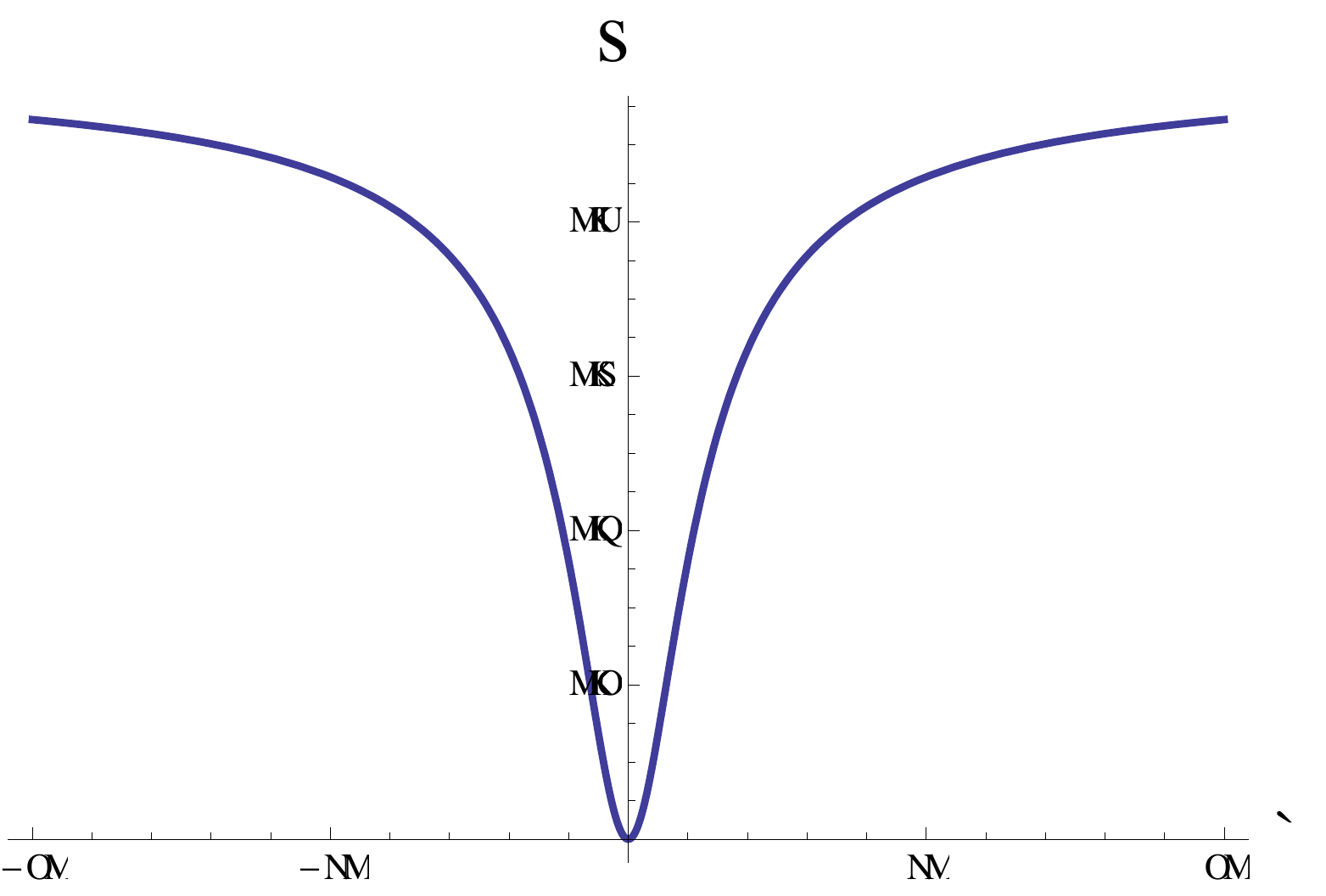}
\caption{The potential of the T-Model as a function of the field $C$. The height of the potential is given in units of $V_{*}$. }
\label{5}
\end{figure}

A particularly interesting   model of this type, which was called the T-Model,  has a potential
\be\label{T}
V = V_{*} \tanh^{2}(\vp/\sqrt 6) \ ,
\ee
which, asymptotically, behaves as $V_{*} (1 -  4 e^{- \sqrt{2/3}\,  \, |\varphi|})$ for $|\vp| \gg 1$. The potential of the T-Model as a function of $C$ is shown in Figure \ref{5}. 
It looks similar to the potential in terms of $\vp$ \cite{Kallosh:2013hoa}. 

The model $R+\gamma R^2$ also can be represented as a member of this class of models, with the inflaton potential $V = V_{*} (1 - e^{-\sqrt{2/3}\, \varphi})^{2} \approx V_{*} (1 - 2 e^{-\sqrt{2/3}\, \varphi}) $ for $\vp \gg 1$ \cite{Kallosh:2013lkr,Kallosh:2013hoa}. 

To analyze this class of models, we will limit ourselves to investigation of the theories (\ref{expansion}), (\ref{T}) at $\vp \gg 1$. 
As long as the coefficient $a_{1}$ in (\ref{expansion}) is not anomalously small, only first two terms matter in our analysis, 
\be\label{expansion2}
V(\varphi) = V_{*} \left[1 - a_{1}  e^{-\sqrt{2/3}\, \varphi} \right] \ . 
\ee
Other terms are exponentially suppressed at large $\vp$; they give a subdominant contribution suppressed by higher powers of $1/N$, where $N =O(60)$ is the number of e-foldings \cite{Kallosh:2013hoa}.
The value of  $V_{*}$ changes the amplitude of perturbations; it is easy to adjust it, so we will concentrate on $n_{s}$ and $r$. According to  \cite{Kallosh:2013hoa}, the inflationary parameters $n_{s}$ and $r$ do not depend on $V_{*}$ and $a_{1}$.\footnote{One can absorb the parameter $a_{1}$ in (\ref{expansion2}) into a redefinition of the field $\vp$ by making a shift under which the kinetic term is invariant. However, the potential of the T-Model is not shift-symmetric, and yet $n_s$ and $r$  do not depend on $a_{1}$ in the first approximation in $1/N$.} Thus, any potential which approaches any positive constant from below as $V_{*}(1-a_{1} e^{-\sqrt{2/3}\, \varphi})$ will belong to the universality class leading to the same predictions for $n_{s}$ and $r$,  in the first approximation in $1/N$ as given in \rf{attractor}.

One should note that these numerical estimates depend on the exact value of $N$, which in turn depends e.g. on the details of the post-inflationary dynamics, including physics of reheating. As a result, the exact results for $n_{s}$ and $r$ may slightly differ from the estimates given above for $N = 60$.

In our new single scalar supergravity with massive vectors/tensors such models with arbitrary $a_1$ are easy to find, as we will show below. However, such models are not exclusive, it is also relatively easy to find supergravity models which depend on different exponents  $e^{-b \vp}$. Models of this type have a long history, starting with~\cite{Goncharov:1983mw}. 
These models with $b \neq  -\sqrt{2/3}$ also lead to $ n_s =1 -2/N$, but the expression for $r$ changes, $r = 8/b^{2}\,N^{2}$, see e.g. \cite{Ellis:2013nxa}. The exponent with $-\sqrt{2/3}$  naturally emerges in the universality class found in \cite{Kallosh:2013hoa}. Let us see that this is not the case here, and that more general exponents can be easily introduced.

We start with the expected potential using the canonical field $\vp$ and identify $C(\vp)$ according to \rf{cphi}. We start with 3 independent constants, $a,b,c$, which will be helpful to compare this model with the geometric version in $C$ variables which we will show below. 
\be
D= c( 1- a \,e^{-b\vp})\, ,  \qquad \rightarrow  \qquad {d D\over d \vp} =  c\,a\,b e^{-b\vp} \ ,
\ee
\be
C(\vp)= - \int d\vp e^{b\vp}/ cab= - e^{b\vp}/ cab^2 \ .
\ee
Thus, 
\be
D(C)= c + {1\over C b^2}\, ,  \qquad D'(C)= - {1\over C^2 b^2} <0 \ .
\ee
Since this corresponds to $J''<0$, the model is a consistent supergravity model for any value of $b$. It will be convenient to set
\be
cab^2=1\, ,  \qquad C(\vp)=  - e^{b\vp}\, ,  \qquad D(C)= c + {1\over C b^2}
\label{Dphi}\ee
Thus we find  a potential
\be
V={g^2\over 2}\Big (c - {1\over  b^2} e^{- b\vp}\Big)^2
\label{genxxxx}
\ee
Note, that the potentials in the family of cosmological attractors found in  \cite{Kallosh:2013hoa} included a broad family of potentials containing higher terms in $e^{- \sqrt{2/3}\, n\, \varphi}$, which did not affect the cosmological predictions in the leading approximation in $1/N$, as long as $a_{1}$ was not anomalously small:  $n_{s} =1 -2/N$, $r = 12/N^{2}$.

By using the same methods as in   \cite{Kallosh:2013hoa}, one can show that 
a similar conclusion should be valid in the case considered above for all sufficiently large values of $b$. Namely, the predictions of the theory with $D= c\left[1 - \sum  a_{n} e^{- b\, n\, \varphi} \right]$ do not depend on $a_{n}$ for a broad range of values of these coefficients:
\be
n_{s} = 1 -2/N\, , \qquad r = 8 /b^2 N^{2}.
\ee
 Indeed, one can show that for $\vp$ corresponding to $N$ e-foldings before the end of inflation, in the leading order in $1/N$,
\be
e^{- b\, \varphi}  = (a_{1} b^{2} N)^{{-1}}.
\ee
As a result, the $n$-th term in the expansion above is $a_{n} (a_{1} b^{2} N)^{{-n}}$, so the higher order terms are suppressed by higher orders in $1/N$. Unless the coefficients $a_{n}$ are anomalously large for $n > 1$, these terms are subdominant for $a_{1} b^{2} N \gg 1$, which is an easy condition to satisfy for $N \sim 60$, $b \geq O(1)$, or $a_{1}$ is anomalously small.

In terms of the function $D(C)$, this implies that theories with $D(C)= c + {1\over C b^2} + O(C^{{-2}})$ are expected to have the same observational predictions, independently of the terms $O(C^{{-2}})$.

The massive vector/tensor supergravity models which we studied in this paper are codified by an arbitrary function, and we have not used any symmetries criteria so far, to pick up only some of the potentials, or classify them. We leave this for future work, we just gave some examples here. It is important that many bosonic models of inflation have found  a simple   supergravity extension in the new framework here.

\subsection{Models with $SU(1,1)$ symmetry}
In \rf{m14}
we described the models defined in~\cite{cfps} where the  \K\, potential gives a manifold $SU(1,1)/U(1)$, and where the gauged symmetry is the translational isometry. It might be useful here 
to  investigate our  new models using some symmetry principles. Consider  more general models with $SU(1,1)$ symmetry:
\be
\Phi(C) = (-C)^{\alpha} e^{\beta C}, \qquad C<0.
\ee
In the language of the de-Higgsed model with $g=0$ and one complex scalar, these  models have a \K\, potential \footnote {Here the \K\,  potential of the Starobinsky model with $\alpha=\beta=1$ has the first term as in no-scale models \cite{Cremmer:1983bf}, however the term  ${3\over 2} \beta (z+\bar z)$ affects the D-term potential and breaks the no-scale structure.}
\beq
K= -3\alpha \ln \left({z+\bar z \over 2}\right)-{3\over 2} \beta (z+\bar z),
\eeq{ms5aa}
and a vanishing superpotential $W=0$. This leads to 
\be
K_{z\bar{z}}\partial z \partial \bar z=3\alpha {\partial z \partial \bar z\over (z+\bar z)^2}.
\ee
This metric corresponds to an $SU(1,1)/U(1)$ symmetric space with constant curvature $R= -2/3\alpha$.
In \rf{m14} we had the case $\alpha=\beta =1$.
Now we have  from $J=3/2 \log \Phi$ that
\be
 J= 3/2 [\alpha \log (-C) + \beta C] \, , \quad J'=3/2(\alpha C^{-1} +\beta)\, ,\quad J''=-(3/2)\alpha C^{-2},
 \ee
 and comparing with \rf{Dphi} we find that
 \be
{3\over 2} \alpha = {1\over b^2}\, , \qquad  {3\over 2} \beta= c\, , \qquad {\alpha\over \beta} =a.
 \ee
So, by setting $C=-\exp(\sqrt{2/3\alpha }\, \vp)$ we find that the potential is
 \beq
V=   {g^2\over 2} (J'(C))^2={9\over 8}g^2  (\alpha/C +\beta)^2,
\eeq{m150}
and
\beq
V={9\over 8}g^2  [\beta- \alpha\exp(\sqrt{2/3\alpha }\, \vp) ]^2,
\eeq{m150'}
in complete agreement with earlier derivation starting from the $\vp$ side of the model. Note that for $\alpha=\beta=1$ we recover the model 
in \cite{cfps}, \cite{Starobinsky:1980te}, which is one of the  attractor models in  \cite{Kallosh:2013hoa}. However, the deviation of $\alpha$ from $1$, i. e. a deviation of $b^2$ from $2/3$
 leads to a more general models with different slow roll parameters, away from the attractor point.  Notice that $\alpha>0$ to have a positive kinetic term for the scalar $z$, while $\beta>0$ for the equation $J'=0$ to have a solution.

 In this setting we find a geometric meaning of the new parameters $\alpha$ and $\beta$, which is difficult to see in the approach which leads to the same final potentials, as given in \rf{genxxxx}. For instance the parameter $\beta$ does not enter in the vector mass $g \sqrt{-J''}$, because it is a \K\, transformation as seen in eq.~(\ref{ms5aa}).
 
Meanwhile, the $\alpha$-parameter defines the curvature of the \K\, manifold, an $SU(1,1)/U(1)$ symmetric space with constant curvature. 
If   primordial gravity waves from inflation will be discovered in the future at the level $\sim 3\cdot 10^{-3}$ in the context of this class of models one would be able to say that the \K\, manifold curvature $R_K$ is given by 
 \be
R_{K}= -K^{-1}_{z\bar z}\partial_z\partial_{\bar z} \log K_{z\bar z} = -2/3\, , \qquad \alpha =1 \ .
 \ee
 But if they are below or above this value, one would be able to say that the discovery of a particular value of $r$ may be viewed as a measurement of a \K\, manifold curvature
 \be
R_{K}= -K^{-1}_{z\bar z}\partial_z\partial_{\bar z} \log K_{z\bar z} = -2/3 \alpha  \ .
 \ee
 This shows that our general model with various predictions may be eventually classified using some geometric symmetry criteria.  
 
\section{Discussion}

For many years there was a certain disconnect between the development of supergravity and cosmology.  The possibility to have inflation in the theories with potentials as simple as $\vp^{2n}$~\cite{Linde:1983gd} without resorting to the cosmological phase transitions required in old and new inflation was a turning point in the development of inflationary theory. It took  17 years until a natural realization of chaotic inflation with a quadratic potential in supergravity was proposed \cite{Kawasaki:2000yn}, and another 10 years to develop a broad class of supergravity models which allowed to have nearly arbitrary inflationary potentials \cite{Kallosh:2010ug}. However, in addition to the inflaton field,  these models also required a Goldstino. Some effort was required to make sure that this field vanishes during inflation. In each particular case this problem was solved by an appropriate choice of the \K\ potential. Also, the presence of extra scalar fields in supergravity inflation could be used, if needed, for generation of non-Gaussian perturbations \cite{Demozzi:2010aj}. However, in the absence of   indication for non-Gaussianity in the Planck data, it would be nice to have as broad class of inflationary potentials as the one developed in  \cite{Kallosh:2010ug}, but without extra moduli fields requiring stabilization. This was one of the goals of the present paper.

In this paper we described a new valley in the supergravity landscape designed to fit all cosmological observations. It is based on a master superconformal gauge-invariant model where the Higgs effect, as well as a de-Higgsing, play an important role. The single  matter multiplet  that is relevant for inflation has to be either vector or linear, but not chiral. The conformal compensators  can be either chiral or linear, thus we have a total of  4 dual models of a new type. Since these are dual models, they are all defined by a single function of one real variable. In the Higgs phase, where we see the relevant cosmological evolution, these superconformal master models 
lead to a cosmological model with a single scalar with an almost arbitrary potential and a massive vector or dual to it massive tensor. 

The final cosmological model of a single scalar is given by  following expression
\be\label{final}
e^{-1} L =  - \frac{1}{2}R - \frac{1}{2}(\partial_\mu \varphi)^2 - \frac{g^2}{2} (D (\vp))^2 \ ,
\ee
where $D (\vp(C))={dJ(C)\over dC} $ and the function $J(C)= {3\over 2} \log \Phi (C) + \rm {const}$ is related in superfield language to the underlying superconformal version of supergravity  by
\be
- [S_0 \bar S_0 \Phi(V)]_D  .
\ee
In components $J$ is related to Jordan frame supergravity as follows
\be
 -{1\over 2 }  \Phi (C) \, R \ ,
\ee
where $C$ is the first component of the real vector multiplet and $R$ is the scalar curvature.
This function  $J(C)$ has to have a negative second derivative
\be
{d^2J(C)\over d^2 C} < 0 \ ,
\ee
corresponding to the \K\, cone restriction.

This restriction implies that the inflaton potential should vanish at its minimum and should grow monotonically away from the minimum in the domain of stability of the theory. 
Apart from that, the class of models developed in our paper is quite general, and many choices can fit easily the area in the $n_s$-$r$ plane favored by Planck data. We gave several examples above. An important advantage of this approach is that in this class of models the  stabilization of moduli is not required, because these models  have only one scalar, the  inflaton field with the action \rf{final}. When   observations   will provide us with more precise values of $n_s$ and $r$, these new models should be capable of fitting them without additional effort. Thus the new approach developed in this paper is an efficient and economical way to relate the early universe cosmology to supersymmetry.

\subsection*{Acknowledgements}
We are grateful to S. Cecotti, V. Mukhanov,  P. Peter, E. Silverstein, and A. Van Proeyen for stimulating discussions.
 The work of RK and AL is supported by the SITP and by  
NSF Grant No. 0756174.   S.F.   is supported by ERC Advanced Investigator Grant n. 226455 {\em Supersymmetry, Quantum Gravity and Gauge Fields (Superfields)}.
M.P. is supported in part by NSF grant PHY-0758032. M.P. would like to thank CERN for its kind hospitality and the ERC Advanced Investigator Grant n. 226455 for support while at CERN. A.L. and R.E. are grateful to the organizers of the School   ``Inflation and CMB Physics'' at Bad Honnef and the School ``Post-Planck Cosmology'' at Les Houches for their hospitality.

\appendix
\section{CFPS Redux}
The SUGRA Lagrangian of the  $ R +\gamma R^2$ theory in the new minimal formulation is given in CFPS~\cite{cfps} as
\bea
{\cal L}&=&-[S_0 \bar{S}_0 \Phi(V_L)]_D + {1\over 2}\gamma [W_\alpha (V_L)W^\alpha(V_L) + h.c. ]_F, \label{mm1}\\ 
&& V_L\equiv \log (L/S_0\bar{S}_0), \qquad \Phi(V_L)=-V_Le^{V_L}, \qquad W_\alpha=\Sigma D_\alpha V_L,
\eea{mm2}
where $\Sigma$ is the chiral projector.

One can re-write the Lagrangian  introducing two real vector Lagrange multiplier $U,V$
\beq
{\cal L}=-[S_0 \bar{S}_0 \Phi(V)]_D +{1\over 2}\gamma [W_\alpha(V) W^\alpha(V) + h.c. ]_F
+U(L-S_0\bar{S}_0e^V ).
\eeq{mm3}
Now, integrating in $U$ we get $V_L=V$, while integrating in $L$ we get $U=\Phi + \bar{\Phi}$, with $\Phi$ chiral. Since $\Phi(V)=-Ve^V$, {\em and only for this special potential} we can redefine $S_0\rightarrow S_0'=S_0 e^{\Phi} $, $V\rightarrow V'=V-\Phi -\bar{\Phi}$. Then the Lagrangian becomes that of a massive vector field:
\beq
{\cal L}=-[S_0 \bar{S}_0 \Phi(V)]_D +{1\over 2}\gamma [W_\alpha(V) W^\alpha(V)+ h.c. ]_F.
\eeq{mm4}

\section{From  Massive Vector to  Massive Tensor in Components}

We derive the component form of the bosonic part of the massive vector supermultiplet theory and its dual massive linear multiplet theory. 

The bosonic part of the massive vector multiplet action, following the conventions of~\cite{vp} is
\beq
{\cal L}(A-\mu,a,H_\mu,C,g)= -{1\over 4} F_{\mu\nu}^2 -{1\over 2} J''^{-1} H_\mu^2
+H_\mu \partial^\mu a + {1\over 2} J''^{-1} (\partial_\mu C)^2 + g H^\mu A_\mu - {g^2\over 2} J'^2.
\eeq{ms6}
Here $H_\mu$ is a Lagrange multiplier. varying with respect to $H_\mu$ we get $H_\mu=J''(\partial_\mu a + g A_\mu)$, which, inserted back into eq.~(\ref{ms6}) gives Lagrangian~(\ref{m10}). The linear multiplet Lagrangian is instead obtained by varying eq.~(\ref{ms6})  with respect to $a$. This gives the constraint $\partial_\mu H^\mu=0$, whose solution is $H_\mu =\epsilon_{\mu\nu\rho\sigma}\partial^\nu B^{\rho\sigma}$. Inserted back into the Lagrangian and after an integrating by part the term $gA_\mu H^\mu$ we obtain
\beq
{\cal L}(A_\mu, B_{\mu\nu}, C, g)= -{1\over 4} F_{\mu\nu}^2 -{1\over 2} J''^{-1} (\partial_{[\mu}B_{\rho\sigma]})^2 + {1\over 2} J''(\partial_\mu C)^2 -g \tilde{F}_{\mu\nu}B^{\mu\nu} -{1\over 2} g^2 J'^2.
\eeq{ms7}
This is the bosonic part of Lagrangian~(\ref{m21}). It can be shown to be independent of $A_\mu$ by replacing $F_{\mu\nu}(A)$ with an unconstrained $F_{\mu\nu}$ and re-writing the two $A_\mu$ dependent terms as follows
\beq
-{1\over 4} F_{\mu\nu}^2 -g \tilde{F}_{\mu\nu}B^{\mu\nu} + V^\mu \partial^\nu \tilde{F}_{\mu\nu}.
\eeq{ms8}
By integrating out $F_{\mu\nu}$ and using the gauge symmetry $B_{\mu\nu} \rightarrow B_{\mu\nu} - {1\over g} F_{\mu\nu}(V)$, eq.~(\ref{ms8}) gives the mass term of the tensor $B_{\mu\nu}$ and so we reproduce Lagrangian~(\ref{m25}).


\end{document}